# What use are Exponential Weights for flexi-Weighted Least Squares Phylogenetic Trees?

Peter J. Waddell[1,2], Xi Tan[2] and Ishita Khan[2]

pwaddell@purdue.edu
tan19@purdue.edu
khan27@purdue.edu

[1]Department of Biological Sciences, Purdue University, West Lafayette, IN 47906, U.S.A.
[2]Department of Computer Science, Purdue University, West Lafayette, IN 47906, U.S.A

.

The method of flexi-Weighted Least Squares on evolutionary trees uses simple polynomial or exponential functions of the evolutionary distance in place of model-based variances. This has the advantage that unexpected deviations from additivity can be modeled in a more flexible way. At present, only polynomial weights have been used. However, a general family of exponential weights is desirable to compare with polynomial weights and to potentially exploit recent insights into fast least squares edge length estimation on trees. Here describe families of weights that are multiplicative on trees, along with measures of fit of data to tree. It is shown that polynomial, but also multiplicative weights can approximate model-based variance of evolutionary distances well. Both models are fitted to evolutionary data from yeast genomes and while the polynomial weights model fits better, the exponential weights model can fit a lot better than ordinary least squares. Iterated least squares is evaluated and is seen to converge quickly and with minimal change in the fit statistics when the data are in the range expected for the useful evolutionary distances and simple Markov models of character change. In summary, both polynomial and exponential weighted least squares work well and justify further investment into developing the fastest possible algorithms for evaluating evolutionary trees.

"Suck it and see." A quotation from Dr Brian McArdle (University of Auckland), frequently given in response to a statistically oriented question.

# 1 Introduction

Distance methods are one of the three main treatments of phylogenetic data used to obtain evolutionary trees (e.g., Swofford et al. 1996). The other two treatments being parsimony-like treatments (including compatibility) and likelihood-based methods (including maximum-likelihood and Bayesian estimators). Distance methods in phylogenetics often involve the collapse of finer grained information (e.g., sequence pattern or allele frequency data) into a matrix of pairwise symmetric distances between all objects (often called taxa in phylogenetic studies). This process of collapsing information into a pairwise distance matrix discards some information for the benefit of a computationally tractable form of data, the distance matrix of order $t_2$ (where t is the number of tips or taxa on the tree). For example, with independent and identically distributed DNA models, information on $4^t$ sequence site pattern frequencies is reduced to $t^2$ distances. How much useful information is discarded with respect to estimating the underlying phylogenetic tree (assuming the data come from a tree) is unclear and there are suggestions that very little is lost in terms of identifying the tree (Roch 2010).

In simulations of DNA sequence evolution, distance methods perform at a similar to slightly lower level of accuracy to likelihood methods, with a major advantage of potentially being much faster to compute. This is no small issue given that searching phylogenetic trees is an NP complete problem that grows as a double exponential each time another taxon is added (Swofford et al. 1996). Thus, strategies of using distance methods to quickly assess trees then pass the best on to full likelihood methods have been described in a variety of guises (e.g., Waddell 1995, Adachi and Hasegawa 1996, Hordijk and Gascuel 2005).

While the discussion in this article is particularly focused on phylogenetic data, it should be remembered that there are other types of data that may fit a tree moderately well and perhaps better than competing visualizations such as multi-dimensional scaling (MDS). This might include expression patterns of cell types in the body (e.g., Ross et al. 2000, Riester et al. 2010), for example, if epigenetic modifications occur in a nearly independent fashion as cells differentiate. For such types of data, least squares distance tree methods are an appropriate exploratory tool (Waddell and Kishino 2000). Indeed, they are the only computationally tractable tools for reconstructing and searching trees with hypothetical ancestors for more than a few thousand tips (Desper and Gascuel 2002). In addition, there are types of data that only come as distances, including DNA hybridization distances and protein structural distances. Ironically, even with phylogenetic sequences there may be an advantage to using alignment free distances over automated alignments (e.g., Hohl et al. 2006)!

If distances are truly additive on the generating tree (for example, counting the number of substitutions on each edge), it is proven that both least squares and minimum evolution type distance methods are consistent (e.g., Denis and Gascuel 2003). Further, least squares methods are robust, for example OLS is still consistent if the errors are heteroscedastic (of unequal variance), with the major drawback being slower convergence to the correct answer (Bulmer 1991). There is an issue that distances have a greater identifiability problem than sequence based likelihood methods. This is because there are fewer distances than site patterns and therefore fewer equations that need to be balanced in order to achieve exactly the same data for two different sets of parameters, including the tree topology (Waddell 1995). They also tend to be less robust then likelihood methods to factors such as unequal site rates (that is, they become inconsistent in identifying the generating tree with less extreme data, Gaut and Lewis 1995, Waddell 1995).

In terms of speed, least squares methods are naively rather slow. However, there are a variety of structures in the operations that can be exploited to reduce ordinary unweighted least squares (OLS) operations from $O(t^4)$ down to $O(t^2)$, with the latter being time optimal. A number of different algorithms have been presented for these operations (e.g., Vach 1989), with those of Bryant and Waddell (1998) offering the flexibility of evaluating binary or non-binary trees. This last form of the operations also offers the fast multiplication of the design matrix that is required to obtain the matrix of residual leverages so useful when studying residuals (e.g., Waddell, Azad, and Khan 2010).

Another extremely useful feature of OLS is the ability to compute the change in edge lengths and score of a tree one nearest neighbor interchange away with a constant computational cost. That allows computing the scores of other trees by successive NNI operations at a constant computational cost of $O(1)$. This allows for algorithms that will find the shortest OLS tree within one partition of the starting tree in $O(t)$ time, within two partitions in $O(t^2)$ time, and within three partitions within $O(t^3)$ operations (Bryant 1996). It is also known that sub-tree pruning and regrafting (SPR) tree rearrangement operations can be done in by a series of $O(t^2)$ NNI operations, which is efficient since there are $O(t^2)$ SPR moves (e.g., the recursion on page 4341 of Hordijk and Gascuel 2005). Similarly, tree bisection and regrafting operations, where the two subtrees may be merged in any form, involves $O(t^3)$ operations and it to can be achieved with $O(t^3)$ NNI operations (Olivier Gascuel, pers. comm.).

A desirable feature of OLS is that if the errors are effectively constant and independent with respect to the size of the distance, then OLS is the maximum likelihood method. Since it is a Gaussian model, then there is a good chance that such a method will have smaller errors and better efficiency than all competing methods under the model. In order to be a truly efficient ML estimator then the solution needs to be limited to the feasible part of the parameter space, which in this case means no negative edge weights. This can be achieved with varying strategies and degrees of success. For speed, if negative edges are reconstructed, their sign can be reversed and the sum of squares taken. This seems to work really well only when the g%SD is small (Waddell, Azad and Khan 2010). A better way is to numerically constrain the solution to positive values using methods such as ellipsoid and interior point algorithms for convex quadratic programming (e.g., Bryant and Waddell 1997, 1998). PAUP*4.0 uses constrained numerical methods and seems to require few cycles to converge (indeed, running PAUP with this constraint is not hugely slower than without it), but the exact algorithms are unpublished (Swofford 2000).

Weighted least squares (WLS) is naively $O(t^5)$ using standard matrix multiplications, but can be computed in $O(t^3)$ for any diagonal vector of weights (Bryant and Waddell 1998). Indeed, the maximum speed of WLS for a single tree is related to the problem of the speed of matrix multiplication. This general problem of multiplying two matrices is readily solvable in $O(t^3)$ time, while complex algorithms allow it to be solved in as little as $O(t^{2.376})$ time (the Coppersmith-Winograd algorithm), although such algorithms are typically only worthwhile for extremely large problems. It is conjectured, with some support, that an $O(t^2)$ algorithm exists, but it too may only be practical for extremely large problems.

A popular method of WLS was introduced by Fitch and Margoliash (1967), where the weights are the distances raised to the power 2 (the FM method). That is, the standard error of the distances is exactly proportional to their size. This was further generalized by Felsenstein (1989) to give a general set of weighted models, a type of flexi-Weighted Least Squares (fWLS), where the weight is $d^P$. Special cases are P = 0 = OLS and P = 2 = FM. The observed distance is often substituted for the expected or tree distance, since with small errors it give similar weights and it avoids the need to iterate to a solution. It comes at the cost of a small loss of efficiency for small errors and at the cost of less predictable statistical behavior (such as the exact size of the expected sum of square under the model) particularly with larger errors. Like OLS, for evolutionary trees WLS should be constrained to the positive part of the parameter space and similar methods are used to achieve this.

A powerful feature of fWLS is that the optimal power of P can be selected by maximum-likelihood (Sanjuán and Wróbel 2005, Waddell, Kishino and Ota 2007). This in turn leads to some freedom from needing to know the exact form of the errors on distances and instead allowing them to be estimated from the data as is typically done with a linear regression model (e.g., Agresti 1990). So far there has been little exploration of how good these approximations are to model based variances, except briefly in Waddell (1995). On the other had, the greatest computational weakness of this method at present is that each tree costs $O(t^3)$ to calculate and there is unfortunately no described speed up for an NNI operation on a precomputed tree.

The most general form of WLS is generalized least squares (GLS) with a full matrix of weights and cross terms, which are usually envisaged as the variances and the covariances of the distances. The iterated maximum likelihood form of this calculation on trees was introduced by Hasegawa et al. (1985). A simplified version without iteration was described independently by Bulmer (1991). Its speed is naively $O(t^6)$ per tree, which is reduced to $O(t^4)$ using the algorithms of Bryant and Waddell (1997, 1998). The structure of covariances in distances from a phylogenetic model is to a large extent described by the tree itself, or more particularly, by the length of the edges that two distances share (e.g., Bulmer 1991). Fortunately, WLS can "eat up" this form of covariance by simply altering the length of the internal edges (Gascuel 1997). How much of the effect of covariances is negated in this way is unclear, but there are claims that WLS does nearly as well as GLS in simulations (Desper and Gascuel 2002).

An interesting method of least squares tree reconstruction is that used by balanced minimum evolution (BME) (Pauplin 2000, reviewed by Gascuel and Steel 2006). It finds edge lengths that minimize the weighted sum of squares, where the weights of the distances are a function of the unweighted tree, for a binary tree they are $2^i$, where $i$ is the number of internal nodes the distance traverses. Thus, its weights are multiplicative across edges or exponential in the total distance measure (which for BME is not the edge length but simply the number of internal internodes crossed). By using a feature of the tree that is invariant to edge lengths, it avoids the need for iteration in order to arrive at multiplicative weights. Its algorithms are fast, being $O(t^2)$ for the first tree and $O(1)$ for an NNI after a full cycle of operations is set up (Hordijk and Gascuel 2005). It has the added advantage that the optimal NNI move about a precomputed tree will always yield a positive edge weight, therefore avoiding the need for a more explicit way to keep the edge lengths non-negative. It's major drawback is that the weights it uses can vary hugely for nearly identical distances or even give short distances much higher implied variances that long distances, depending on the form of the tree being considered (e.g., Waddell, Azad and Khan 2010). A second draw back is that it optimality criterion is not weighted least squares, which would be maximum likelihood under its model of edge length errors, but the less efficient minimum evolution criterion, of simply summing up the WLS edge weights.

Recently it has been shown that the optimal least squares edge weights for any form of an a priori multiplicative weight may be estimated in $O(t^3)$ time using Lagrange multipliers (Mihaescu and Pachter 2008). This hints at future possibilities for WLS, but much work remains to be done. For example, there is no described means of performing an NNI in less than $O(t^3)$ time, nor is there any way of constraining the solution to the non-negative part of the parameter space. Even if there is a fast way of recalculating the edge lengths after an NNI, this only leads to a faster from of minimum evolution, without a fast algorithm for updating the sum of squares (which is naively $O(t^2)$). Further, the use of any multiplicative weights that are functions of edge lengths requires iterative least squares.

Perhaps most significantly, there is no case yet made that multiplicative models could be generally useful in the way that fWLS polynomial weight models seem to be (Waddell and Azad 2009). This requires the study of how the signal-to-noise ratio changes with respect to the evolutionary distance (Waddell 1995). Without this clarified there is little incentive to invest time into attempting to solve computational issues. There is also no firm intuitive foundation for predicting their relevance to different types of data. In this article we explore the properties of a general form of fWLS multiplicative model using exponential weights, contrast it to the polynomial weights model, and ascertain its usefulness for phylogenetics and related fields such as trees of gene expression data.

## 2 Materials and Methods

The yeast alignment of 107 protein-coding genes from 8 species with a total length of 127,026 nucleotide sites is from Rokas et al. (2003). These were translated to amino acid sequences using the universal nuclear code. Amino acid distances we calculated using the using ML + Γ distance of Waddell and Steel (1997) as described in Waddell and Azad (2009).

Fit of trees to data was performed using PAUP* (Swofford 2000), FITCH and recompiled FITCH code for exponential weights (Felsenstein 1989), and MATLAB code.

Parametric "replicate" sequence data sets were generated using Seq-gen (Rambaut and Grassly 1997). C programs and PERL scripts were used to automate and integrate tasks, including the resampling of residuals that were then feed to other programs to calculate distances and then to PAUP for tree search. Simple graphics and analyses were made using Microsoft Excel.

## 3 Results

The first results section 3.1 describes general families of exponential weights that are multiplicative across edges of a tree. Section 3.2 shows how the log likelihood and the g%SD fit measure are calculated for fWLS with exponential weights. The third results section involves polynomial and exponential fWLS compared to model based estimates of the variance, with and without unequal site rates. Section 3.4 evaluates both models fitting trees to yeast evolutionary distances. The fifth section considers the vexing question of iteration, which is required to exploit current fast tree algorithms with exponential weights. Finally section 3.6 compares the fit of exponential and polynomial weighted trees on the much larger gene expression profile data set of Ross et al. (2000).

### 3.1 A General Model of Multiplicative Weights

The sum of squares of the standard or polynomial weights model of flexi-Weighted Least Squares (fWLS) is of the form

$$\sum_{i=1}^{N} \frac{(d_{obs_i} - d_{exp_i})^2}{d_{exp_i}^P} = \sum_{i=1}^{N} \frac{(d_{obs_i} - d_{exp_i})^2}{\ln[e^{Pd_{exp_i}}]}, \qquad \text{(eq 1)}$$

e.g., Felsenstein (1989). Here we evaluate the behavior and performance of a novel multiplicative or exponential model of fWLS where the sum of squares is of the form

$$\sum_{i=1}^{N} \frac{(d_{obs_i} - d_{exp_i})^2}{e^{Pd_{exp_i}}}. \qquad \text{(eq 2)}$$

Note that at P = P' = 0 both models reduce to the ordinary unweighted least squares (OLS) estimator. Otherwise, their behavior could be very different for different values of P or P', depending on the scale of the distances. Thus, when we compare them we tend to use the somewhat scale-invariant parameter P, and equate P' to it by either matching some more fundamental characteristic, such as the scaled range, coefficient of variation (that is the quantity of the standard deviation of the weights divided by the mean weight) or log ratio of the weights yielded by respective functions on the data at hand. For convenience, we also use the approximation of using $d_{obs}$ in place of $d_{exp}$ in some calculations below. This is computationally much more tractable, will yield minor differences in terms of percentage error, and should not interfere with our purpose here of determining how well the multiplicative weights either match real patterns of error in the data or how well they approximate the polynomial fWLS model. This assumption is check further below.

As an aside, the model underlying balanced minimum evolution (BME, Pauplin 2000) can also be generalized so that the sum of squares minimized in estimating the edge lengths is of the form, $\sum_{i=1}^{N} \frac{(d_{obs_i} - d_{exp_i})^2}{2^{P'x_i}}$ where, x is the number of internodes a distance traverses on the tree being evaluated. Here, if P' = 0 then the function is OLS and if P' = 2 the model is BME. If fast algorithms could be found for this form of weight then some of the problems with BME weights being too extreme on certain types of weighted tree could be mitigated (e.g., Waddell, Azad and Khan 2010).

## 3.2 Fit Criteria for the Multiplicative Model

The general form of the geometric mean percentage fit equations in Waddell, Kishino and Ota (2007) and Waddell and Azad (2009) is given in Waddell, Azad and Khan (2010). If the model of the variance is $\sigma_i^2 = \sigma^2 w_i = c w_i$, then we have the general scale-free fit monotonic with the likelihood being,

$$g\%SD = \left(\prod_{i=1}^{N} d_{obs_i}\right)^{-1/N} \left(\prod_{i=1}^{N} w_i\right)^{\frac{1}{2N}} \left(\frac{1}{N}\sum_{i=1}^{N}\frac{(d_{obs_i} - \hat{d}_{\exp_i})^2}{w_i}\right)^{0.5} \times 100\% \qquad \text{(eq 3)}$$

which, for the exponential weight model in equation 2 becomes,

$$g\%SD = \left(\prod_{i=1}^{N} d_{obs_i}\right)^{-1/N} \left(\prod_{i=1}^{N}\exp(P'd_i)\right)^{\frac{1}{2N}} \left(\frac{1}{N}\sum_{i=1}^{N}\frac{(d_{obs_i} - \hat{d}_{\exp_i})^2}{\exp(P'd_i)}\right)^{0.5} \times 100\% \qquad \text{(eq 4)}$$

This is the likelihood inspired version of the percentage fit statistic. However, since these models fit into the framework of general linear models, these statistics can also be based on unbiased estimators of the variance. In this case replace the normalization term $1/N$ immediately before the sum of squares with the term $1/(N-k)$, where $N$ is the number of pieces of information (the number of unique distances) while $k$ is the number of free parameters estimated by the model. This yields,

$$g\%SD = \left(\prod_{i=1}^{N} d_{obs_i}\right)^{-1/N} \left(\prod_{i=1}^{N}\exp(P'd_i)\right)^{\frac{1}{2N}} \left(\frac{1}{N-k}\sum_{i=1}^{N}\frac{(d_{obs_i} - \hat{d}_{\exp_i})^2}{\exp(P'd_i)}\right)^{0.5} \times 100\% \text{ ,} \qquad \text{(eq 5)}$$

which is the form of the fit reported below. Note, this property of unbiasedness only strictly holds when the generating tree is known and it is allowed to have negative edge lengths. Otherwise, the true expected values of the variance deviate due to tree selection, constraint of edge lengths to be non-negative and the constraint that $d_{exp}$ be non-negative (the latter being a weaker condition than the former).

Note, it is easy to interconvert the g%SD calculated with different normalizations of the empirical sum of squares. For example, to convert the value estimated with $1/N$ to that estimated with $1/(N - k)$ simply multiply by square root of $N/(N - k)$. This fit measure has recently been added to PAUP*4.0 a114 (Swofford 2000) and there the normalization used is $1/(N-1)$.

## 3.3 How multiplicative are Markov model variances?

A key piece of information to understanding how fWLS with either polynomial or exponential weights works is to understand the relationship of these power functions to model data. Work on this is found in Waddell (1995) where the statistical basis of the Hadamard conjugation and its relationship to maximum likelihood and tree space is explored. Figure 1 is reproduced here from there, and it exploites the key concept of the signal to noise ratio of the distances, which is closely related to the inverse of the percentage fit statistics (or more properly, the proportional fit statistics, without the multiplication by 100%).

Examining figure 1a, notice that when the signal to noise curve is flat, then the signal to noise ratio is constant, which fits the assumptions of the Fitch Margoliash (1967) model, that is, the polynomial weights model with P = 2 (Waddell 1995). Notice also that the more extreme the shape of a gamma distribution of site rates, the closer to flat a larger portion of the curve becomes. That is, a gamma distribution of site rates causes the peak signal to noise ratio to occur at shorter evolutionary distances, it decreases the height of the peak, and after the peak it dies away more slowly. Thus, in this situation the error on the distances comes closer to fitting the expectations of P = 2 over a broad range of evolutionary distances.

Figure 1b shows that increasing the number of character states raises the signal to noise ratio and also results in a broader region of nearly constant signal to noise ratio. The observations in figure 1a and 1b agree with using polynomial weights to fit the GTR G amino acid distances for yeast data from Waddell and Azad (2009). There, using the polynomial weights model there was a well-defined minimum at P = 2.3.

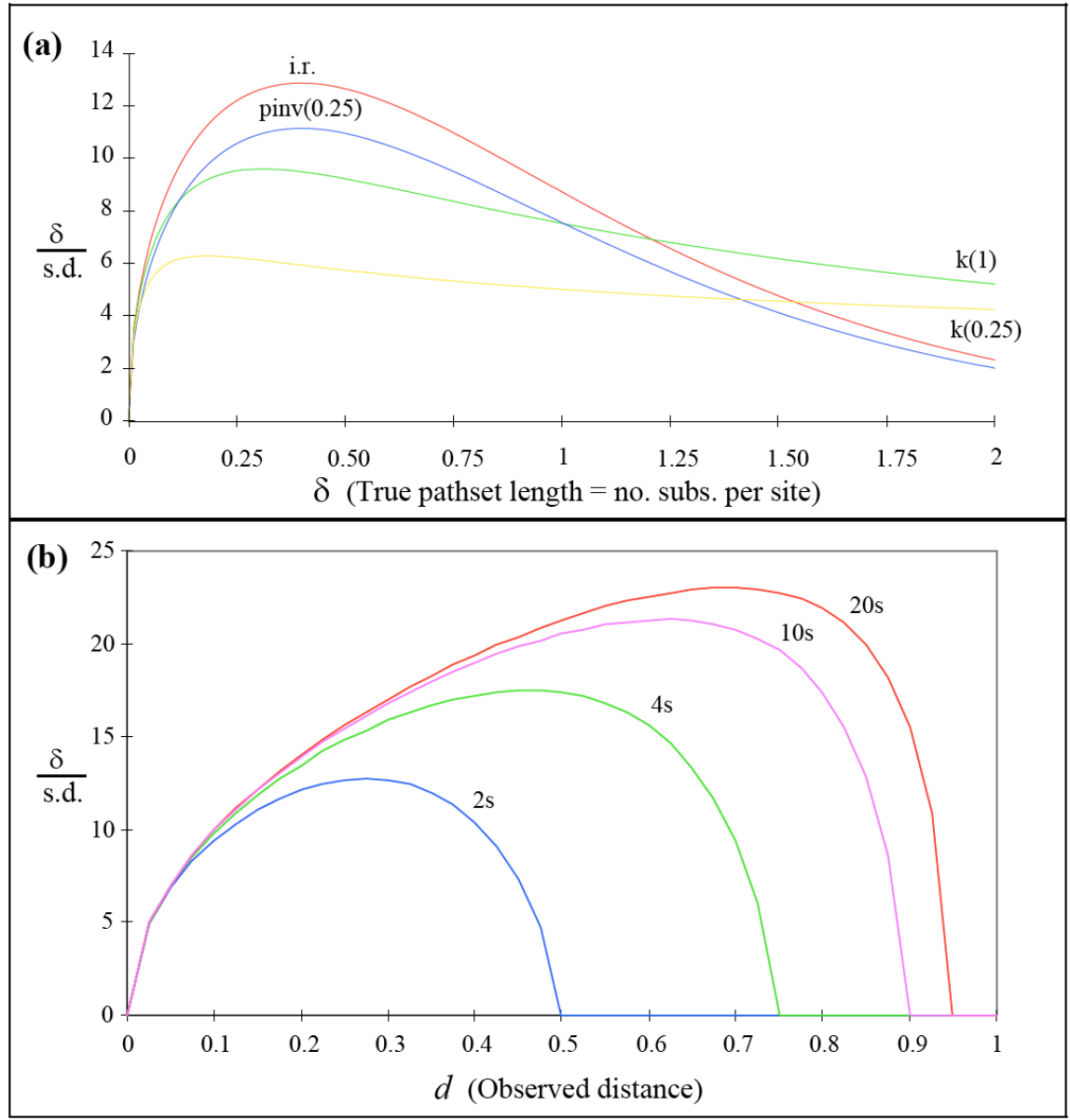

Figure 1 (a) A plot of the signal to noise ratio for various site-rate distributions and a 2-state evolutionary distance with a sequence length of *n* = 1000 sites. The signal to noise ratio is measured by the true evolutionary distance divided by its standard deviation (estimated using a delta-method approximation of the variance, Waddell et al. 1994, Waddell, Penny and Moore 1997). The basic substitution model is that of Cavender-Felsenstein, which is the 2-state equivalent of the Jukes-Cantor model (Swofford et al. 1996). This may be called a simple Poisson-based evolutionary distance. The red line marked i.r. indicates all sites have identical rates; the blue line marked $p_{inv}$ is all sites have identical rates except 25% that cannot change; the green line indicates that site rates have a gamma distribution with shape parameter *k* = 1; lastly, the yellow line indicates that the site rates have a more extreme shape parameter of 0.25 (e.g., Waddell, Penny and Moore 1997). (b) Extending the signal to noise plot of an identical rate Poisson evolutionary distance to a range of character state numbers (namely, 2, 4, e.g. DNA, 10 and 20, e.g. amino acids). Note that the $x$-axis uses the Hamming or observed distance to compact this axis. The dominant effect of more states is to produce a larger peak at larger values, but with a broader region of near flat signal to noise ratios (with the last effect exacerbated if site rates follow a gamma distribution with shape less than 1). Reproduced with permission from Waddell (1995), page 210, figure 4.7.

Next we test the suggestion of Mihaescu and Pachter (2008) that the delta method variance of the Jukes Cantor substitution model (e.g., Swofford et al. 1996) was a good match to a multiplicative model. It is apparent in comparing their variance formula 27 with that of Kimura and Ota (1972),

$$Var(D) \approx \frac{p(1-p)}{[(1-4p/3)^2 n]},\tag{6}$$

who first derived a delta method approximation to the Jukes-Cantor model, that there are issues (here Var(D) is the variance of the evolutionary distance, $n$ is the number of sequence sites and $p$ is the proportional mismatch, observed, or Hamming distance). The formula 27 equation gives estimates of the variance are approximately 3 times larger than those of Kimura and Ota. Thus, the signal to noise ratio is nearly two times too small, as seen in figure 2a. Simulations have shown that delta method approximations of the form derived by Kimura and Ota (1972) can be very accurate. For example, with the 2-state Cavender-Farris model, Waddell et al. (1994) showed that the delta method was an accurate predictor of the true variance in the range of distances that matter most in phylogenetics (more so than the mean of bootstrap replicates). By modifying the formula of Mihaescu and Pachter to,

$$Var(D) \approx \frac{1}{16n}[3e^{8\delta/3} + 2e^{4\delta/3} - 3)]\tag{7}$$

then the asymptotic values of the signal to noise ratio converge more quickly and the overall agreement is much better as seen in figure 1a. However, even then equation 7 does not converge to the binomial variance as it should for small evolutionary distances, $\delta$. Thus, notice that the exponential approximation of equation 7 is not always a close one, especially at lower values where the formula of Kimura and Ota converges (as it should) to the simple binomial variance of $p(1-p)/n$.

Next, compare polynomial and multiplicative weights to the delta method variance of Kimura and Ota in the general range in which a Jukes-Cantor distance is most frequently used, that is $\delta$ from zero to 0.5. If the measure of correspondence of weights to variance is the Pearson correlation coefficient, $\rho_{xy}$ (when forcing the variance at zero distance to be zero, as it must be), then the polynomial weights had the best correspondence with P = 1.5 yielding $\rho_{xy}$ = 0.9994 while for multiplicative weights it was P' = 3.5 and $\rho_{xy}$ = 0.9931. However, since the correlation coefficient forced to go through the origin is particularly influenced by the colinearity of the weights to the variance for the larger distances, a graphical approach can be more informative. Figure 2b shows the closest visual match arrived at for polynomial and multiplicative weights against the delta-method variance. Clearly, the polynomial weights do a markedly better job of mimicking the true variance in this critical range of evolutionary distances. The multiplicative weight is not as good, but it does capture some of what is going on. The multiplicative weights do reasonable job of mimicking equation 7, which suggests that it may also be able to mimic moderately well functions that are a mixture of exponential functions.

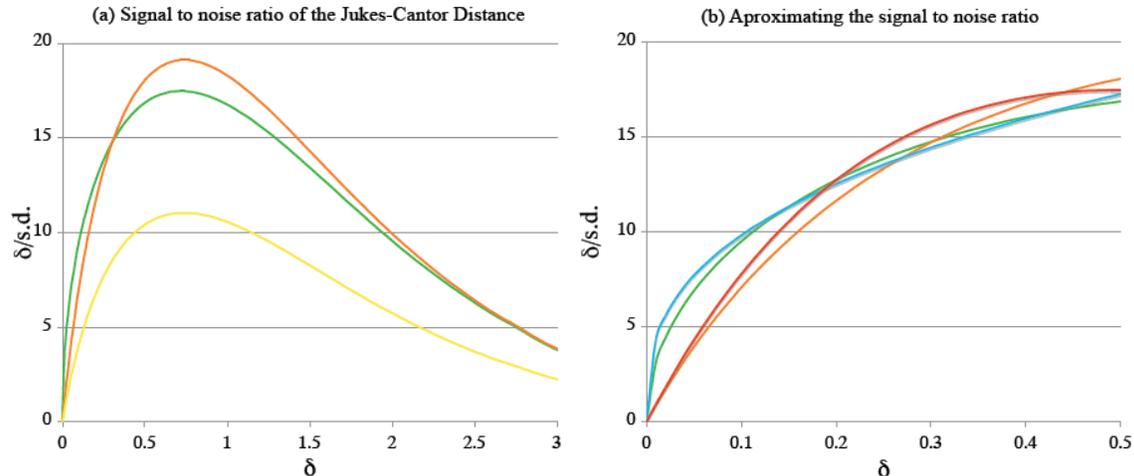

Figure 2 (a) A plot of the signal to noise ratio of the Jukes-Cantor distance. Here, δ is the true evolutionary distance. The green line is produced using the delta-method approximation of the variance from Kimura and Ota (1972), which has been validated in simulations (e.g., Waddell et al. 1994), the yellow line is the delta-method approximation presented in equation 27 by Mihaescu and Pachter (2008), while the orange line is using equation 7 above. (b) Looking at how different forms of weights can approximate the variance (green line) of the Jukes Cantor model in the critical region of evolutionary distances up to 0.5 substitutions per site. The blue line shows polynomial weights of the form $c\delta^P$ (P = 1.3) and the red line are exponential weights of the form $c'exp(P'\delta)$ (P'= 4). The orange line follows equation 7.

### 3.4 Direct comparison of polynomial and multiplicative weights on sequence data

Here the two main models, fWLS with polynomial and fWLS' with exponential weights are compared on yeast amino acid GTR distance data from Waddell and Azad (2009) based on the sequence alignments of Rokas et al. (2003). Figure 3 shows the results. Both models have their minima at similar points with that of the polynomial model being at P = 2.3 and g%SD = 1.0802. The curve for polynomial weights is symmetric and the weights have lead to a markedly better fit than the unweighted least squares model (P = 0). The fit of the multiplicative model is somewhat worse with a minimum at P' = 3.25 and g%SD = 1.4123, but still a marked improvement over the unweighted least squares model. Unlike the polynomial weights the curve for the multiplicative model heads towards a second minimum for negative weights. This type of behavior is data dependent and occurs in some data sets with polynomial weights. It might be interpreted as a warning that the model weights and the residuals do not uniquely coincide, indicating that there remains some disagreement on the basic form of the model.

In order to better compare these two models we plotted both curves on a common axis. A number of forms of the x-axis are considered, including the coefficient of variation of the weights (with a change of sign below P = 0). Shown in figure 4 is the range of the weights (weight of the largest distance minus that of the smallest) divided by the mean of all weights. On this axis the minimum of the two models appears to coincide even more closely, that is, when the range of the weights divided by the mean weight is about 4. It is also more apparent what has occurred with negative weights. The exponential function is losing potency compared to the polynomial function due to the scale of the distance and we have completed prematurely. Indeed, if the multiplicative function is followed to larger negative values it soon hits a clear second minima then, like the polynomial function, gets progressively worse.

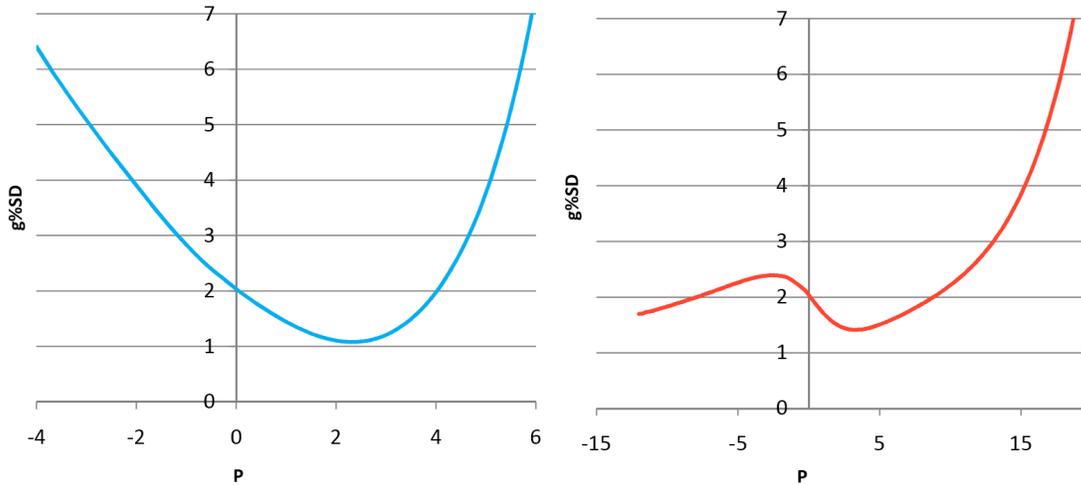

Figure 3 (a) A plot of the fit (g%SD) of the fWLS polynomial weight or standard model against the free parameter P, such that the weights are $d_{obs}^P$ for the yeast GTR AA distance data of Waddell and Azad (2009). It has a minimum at P = 2.3, g%SD = 1.0802. (b) A plot of the fit (g%SD) of the fWLS exponential weight or multiplicative model against the free parameter P', such that the weights are exp(P'$d_{obs}$). The minimum occurs at P' = 3.25 and g%SD = 1.4123.

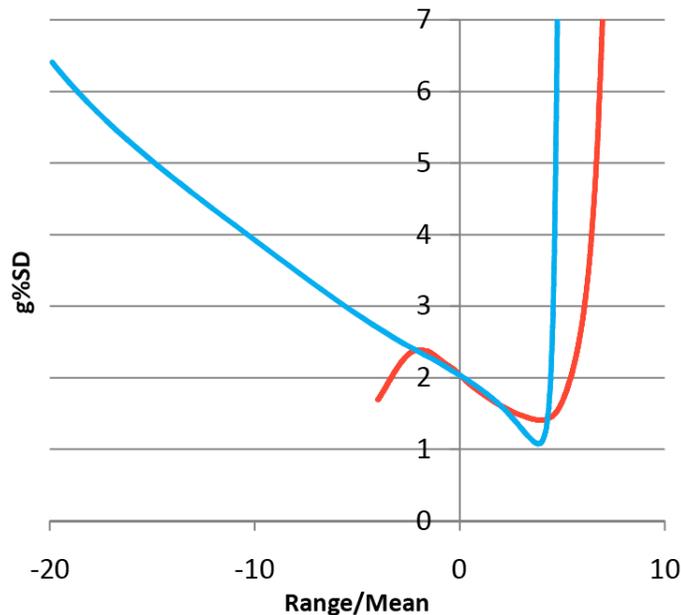

Figure 4. A plot of the fit (g%SD) of the fWLS polynomial model (blue) and the fWLS multiplicative model (red) when the x-axis records the range of the weights divided by their mean. For weights less than zero, the sign of the range is changed, indicating the larger distances now have smaller weights (that is, the range is defined as the weight of the largest distance minus the weight of the smallest distance).

### 3.5 Iterated flexi-weighted least squares

In this section we consider how iterated least squares, differs from that of uniterated least squares. The two reasons to do this are firstly statistical; using the expected distances should be slightly more discriminatory (statistically efficient) than using the observed distances (e.g., Agresti 1990). Secondly, are computational reasons. If fast algorithms for multiplicative weights based on edge lengths through a tree are to be useful, then iteration is required as discussed earlier. A series of iterative results for the yeast distance data and multiplicative weights are shown in figure 3.

Figure 5a clearly shows the extent of the second dip in the fit curve for exponential weights. The minimum in the negative range of P' is slightly worse for uniterated and slightly better for iterated weights than is the minimum in the range of positive P. Fortunately, this minimum is well away from the region we would usually consider has any relation to the variance on evolutionary distances of a meaningful nature.

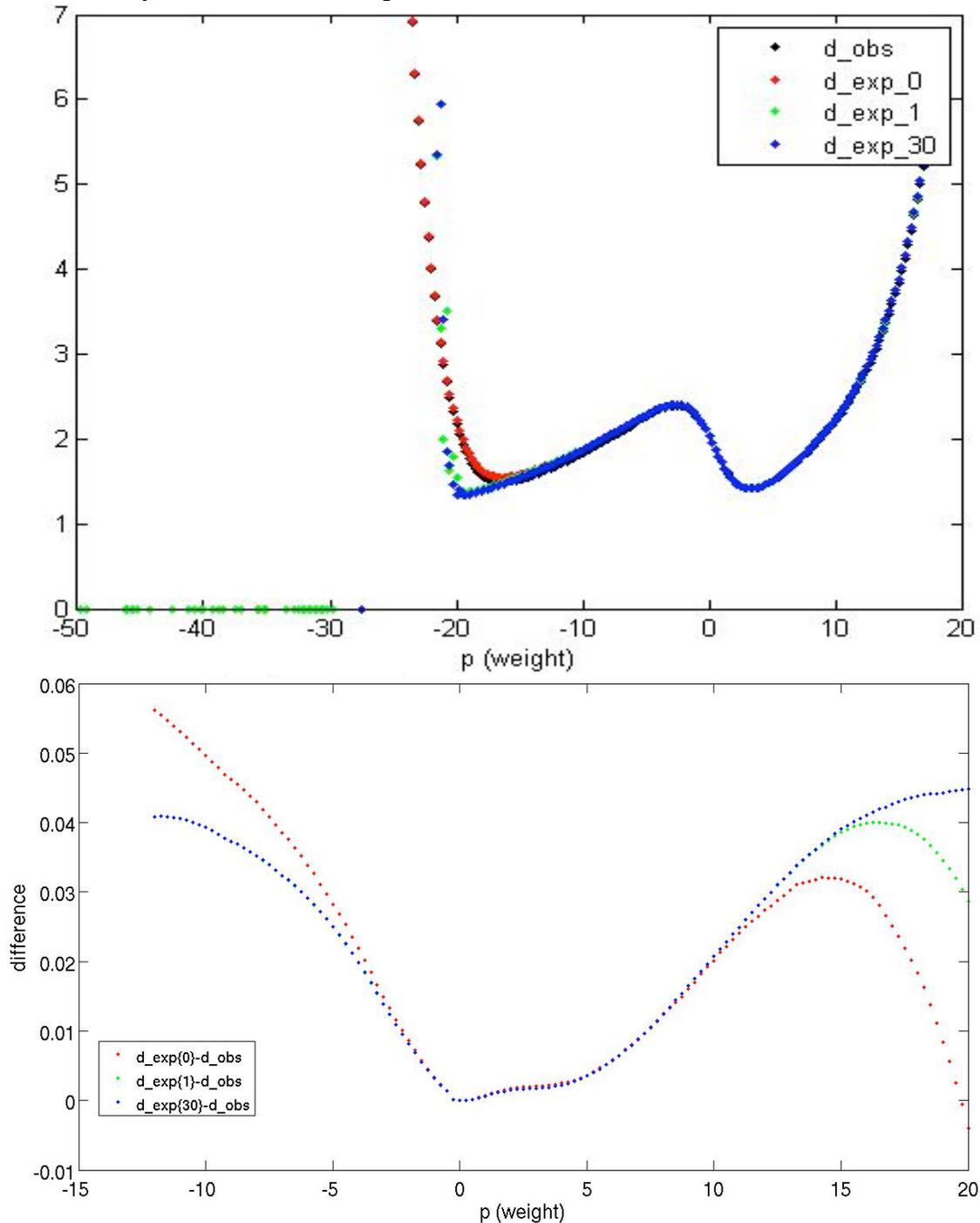

Figure 5. Plots of the fit (g%SD) of the fWLS multiplicative model. (a) A plot of parameter P' versus the g%SD for iterated least squares on the yeast distance data. The uniterated least squares weights are $d_{obs}$, $d_{exp}(0)$ is the same as previously but using the model distances in the evaluation of the g%SD, $d_{exp}(1)$ has updated weights after one cycle, while $d_{exp}(30)$ has updated weights after 30 cycles (on this data,

convergence was nearly always completed within 10 cycles). (b) As previously but the three curves are the differences of g%SD $d_{exp}$ after different numbers of cycles to that with $d_{obs}$ plotted against parameter P'.

After the optimal edge lengths are calculated for weights of the form exp(P'$d_{obs}$) only a few cycles (less than ten in the cases examined) were required for complete convergence to the precision of the machine. Another encouraging result was that the g%SD changed very little with iteration as seen in figure 3a at values in the range of desired fit, that is with a log ratio of weights between 0 and perhaps 5. The actual differences due to iteration are indicated more clearly in figure 3b. Here it is seen that the difference of iterated and uniterated least squares is only really noticeable away from P = 0, where both models reduce to unweighted least squares. The similarity of the red and the blue lines between about P' = -5 to P' = 12 shows the difference is mostly on the zero0th cycle by substituting in the edge lengths calculated with exp(P'$d_{obs}$) with the edge lengths arrived at with these weights. The green line shows that after one more iteration, the final result is only visibly different from iteration to convergence for P' > 15.

Comparable results are shown for the polynomial weights in figure 6. Here, some differences in the g%SD fit are more apparent with iteration, but they are only apparent away from both P = 0 and the minimum at P = 2.3.

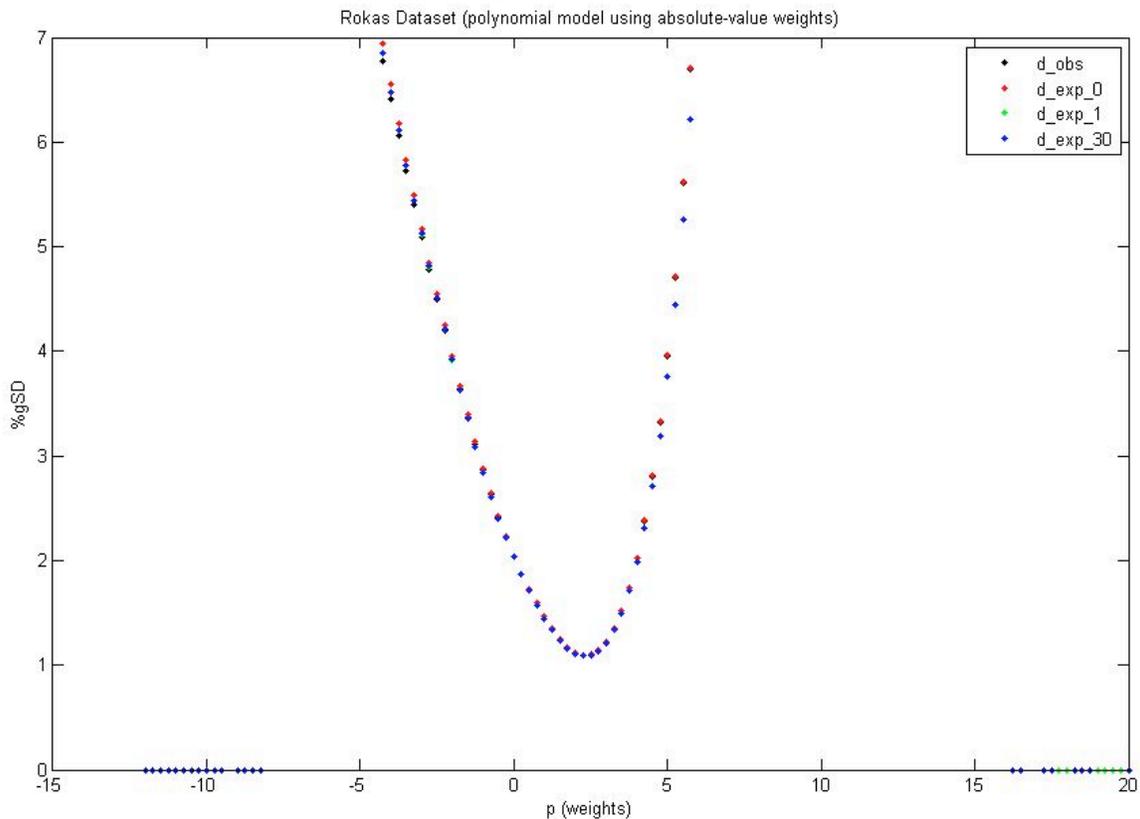

Figure 6. A plot of the fit (g%SD) of the fWLS polynomial weights against parameter P for the yeast amino acid GTR distances. (a) The uniterated least squares weights are $d_{obs}$, then $d_{exp}$(0) which the same as previously but using the model distances, $d_{exp}$(1) with updated weights after one cycle, $d_{exp}$(30) with weights after 30 cycles (on this data, convergence was nearly always completed within 10 cycles). (b) As previously but the three curves are the differences of g%SD $d_{exp}$ after different numbers of cycles to that with $d_{obs}$.

### 3.5 The gene expression data from Ross et al. 2000

Figure 7 shows the fit of polynomial and exponential weights to the gene-expression data of Ross et al. (2000) on two different common axes. The same data are evaluated with residual resampling in Waddell, Azad and Khan (2010). This was a seminal data set in that it revealed clear clustering of caner cell lines based on their tissues of origin when first publish a decade ago. It also facilitated the detection of "cryptic" cancer types, which histologically were undistinguished, but which showed very different expression profiles. The original authors used a combination of hierarchical clustering (via the ultrametric UPGMA tree inference algorithm) and Principal Components Analysis (PCA) to visualize the results. Here we see a surprising result, in that the fit of these data sets to a tree is markedly better than might be expected by chance, with a g%SD of 7.590 at P = 0 to -0.3 (so basically, ordinary least squares fitting). While this g%SD is not nearly as "clean" as a typical gene tree (which is often less than 2% for mammalian genes longer than 1 kb) it compares very favorably in its tree-likeness with nearly all morphological evolutionary data sets studied so far (e.g. those at morphbank), which have a g%SD typically in the range of 10 to > 30 (unpublished results)!

That the minimum is near zero for both models, implies that the size of errors on fitting to a tree remain pretty much the same irrespective of the difference in the gene expression profile. This also indicates that methods such as balanced minimum evolution (BME), which are based on reconstructions using highly different weights, may struggle in analyzing such data (Waddell, Azad and Khan, 2010). Notice also that the curve is fairly flat across the whole range of P and P' values which can be very different. Analyzing this data at P = 0 rather than the true minimum at P = -0.1 or -0.2 is a minor issue in terms of fit and it takes advantage of the fast algorithms for OLS (Bryant and Waddell 1998) implemented in PAUP* (Swofford 2000). These are order ($t^2$) per tree rather than order ($t^3$) per weighted least squares tree as currently programmed.

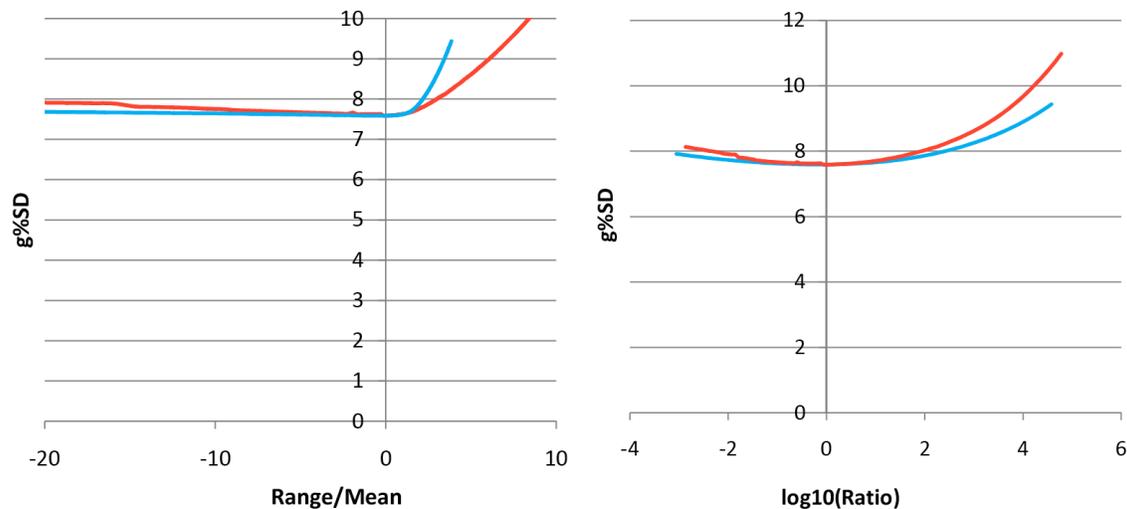

Figure 7. (a) A plot of the fit (g%SD) of the fWLS polynomial model (blue) and the fWLS multiplicative model (red). The x-axis records the range of the weights divided by their mean. For weights less than zero, the sign of the range is changed, indicating the larger distances now have smaller weights (that is, the range is defined as the weight of the largest distance minus the weight of the smallest distance). The minima occur at P = 0 to -0.3 and g%SD, plus P' = -0.1 and a g%SD of 7.590. (b) The same data but the x-axis is represented as the log to base 10 of the ratio (weight[max $d_{obs}$] / weight[ min $d_{obs}$]) which is the log of the ratio of the weights of the largest to those of the smallest distance in the data.

## 4 Discussion

Overall, the results are encouraging for considering further development of the exponential or multiplicative weights models for fWLS phylogenetic trees. While the fit of a polynomial model seems to better approximate the variances under simple Markov models of evolution (Swofford et al. 1996), the flexible multiplicative weights introduced in this article can do a reasonable job, and potentially, enough of an improvement over assuming all distances have the same errors to encourage their wider use.

With real data, we also see that the multiplicative model tends to share a lot in common with the polynomial model, particularly when the two are plotted on a common x-axis comprising the log of the ratio of the weight of the largest distance to the weight of the smallest distance. On the Ross et al. (2000) data, for example, even the worst multiplicative weights explored (with extreme weights on different distances) gave much better fits than the unusual multiplicative node weights of BME (Waddell, Azad and Khan 2010).

The polynomial model with weight P = 2 is seen to be near a potential sweet spot for analyzing DNA and amino acid sequence with large site rate variability following a gamma model (e.g., Waddell 1995, Waddell, Penny and Moore 1997). It will be interesting to see if this prediction from theory pans out further with real data sets. Unfortunately, as far as is known the value of P = 2 does not offer any special computational advantage over other values of P or WLS using variances from the model.

It was seen here that iterated weighted least squares can converge fairly quickly for both multiplicative and polynomial fWLS. This will hopefully encourage further research into iterative algorithms. Indeed, there may be practically fast iterative algorithms for tree-like data (as many real data sets will be) that combine iteration to keep edge lengths non-negative, to use expected rather than observed distances in weighting and, for multiplicative weights, to realize potential computational benefits over all types of weighted least squares.

The advantage of flexi-Weighted Least Squares remains with its ability to adjust to the real errors encountered in the data and not just those hypothesized by simple models. In this way they may bring some of the sensibilities of typical regression analyses that are sadly missing in phylogenetics into this important but still underdeveloped field. As such, both the polynomial weights and the multiplicative weights models deserve further exploration into their computational speed up, particularly for rearrangements about the first tree and for iteration to either achieve realistic multiplicative weights and/or to constrain edge lengths and/or use expected distances for weights. In conclusion, we have sucked on the question of what use if any are multiplicative weighted least squares and discovered that both they and polynomial weights hold potential for algorithmic extensions useful in the real world of the analysis of tree-like data using general linear models.

## Acknowledgements

This work was supported by NIH grant 5R01LM008626 to PJW. Thanks to David Bryant, Joe Felsenstein, Olivier Gascuel, Hiro Kishino, Radu Mihaescu and Dave Swofford for helpful discussions.

## Author contributions

PJW originated the research, developed methods, gathered data, interpreted analyses, prepared figures and wrote the manuscript. XT and IK implemented methods in C and PERL, ran analyses, prepared figures, interpreted analyses and commented on the manuscript.